\documentclass{nature}
\usepackage{amssymb}
\usepackage{aas_macros}
\usepackage{graphicx}
\usepackage{pdflscape}
\usepackage{hyperref}
\usepackage{txfonts}
\usepackage{textcomp}
\usepackage{threeparttable}
\usepackage{marvosym}

\usepackage{amssymb,amsmath}
\usepackage{xcolor}
\usepackage{ulem}
\usepackage{lineno}
\usepackage{pdfpages}
\usepackage[caption = false]{subfig}
\usepackage{caption}
\makeatletter
\usepackage[symbol]{footmisc}
\newcommand{\EXTTAB}[1] {supplementary table ~\ref{#1}}
\usepackage{longtable}

\let\saved@includegraphics\includegraphics
\AtBeginDocument{\let\includegraphics\saved@includegraphics}
\renewenvironment*{figure}{\@float{figure}}{\end@float}
\makeatother

\newcommand{\lsim}{{\;\raise0.3ex\hbox{$<$\kern-0.75em\raise-1.1ex\hbox{$\sim$}}\;}}
\newcommand{\gsim}{{\;\raise0.3ex\hbox{$>$\kern-0.75em\raise-1.1ex\hbox{$\sim$}}\;}}

\definecolor{dkblue}{RGB}{54, 86, 169}

\title{A bimodal burst energy distribution of a repeating fast radio burst source}

\author{
D. Li$^{1,2}$\thanks{Email:dili@nao.cas.cn, \href{https://orcid.org/0000-0003-3010-7661}{orcid.org/0000-0003-3010-7661}}{\includegraphics[scale=0.08]{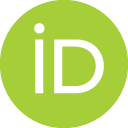}}\textsuperscript{\Letter}\footnotemark[1],
P. Wang$^{1}$\href{https://orcid.org/0000-0002-3386-7159}{\includegraphics[scale=0.08]{ORCIDiD.png}}\footnotemark[1],
W. W. Zhu$^{1}$\href{https://orcid.org/0000-0001-5105-4058}{\includegraphics[scale=0.08]{ORCIDiD.png}}\footnotemark[1],
B. Zhang$^{3}$\thanks{Email: zhang@physics.unlv.edu, \href{https://orcid.org/0000-0002-9725-2524}{orcid.org/0000-0002-9725-2524}}{\includegraphics[scale=0.08]{ORCIDiD.png}}\textsuperscript{\Letter},
X. X. Zhang$^{1}$\href{https://orcid.org/0000-0001-9587-631X}{\includegraphics[scale=0.08]{ORCIDiD.png}},
R. Duan$^{1}$\href{https://orcid.org/0000-0002-9260-1641}{\includegraphics[scale=0.08]{ORCIDiD.png}},
Y. K. Zhang$^{2,1}$\href{https://orcid.org/0000-0002-8744-3546}{\includegraphics[scale=0.08]{ORCIDiD.png}},
Y. Feng$^{2,1,4}$\href{https://orcid.org/0000-0002-0475-7479}{\includegraphics[scale=0.08]{ORCIDiD.png}},
N. Y. Tang$^{5,1}$\href{https://orcid.org/0000-0002-2169-0472}{\includegraphics[scale=0.08]{ORCIDiD.png}},
S. Chatterjee$^6$\href{https://orcid.org/0000-0002-2878-1502}{\includegraphics[scale=0.08]{ORCIDiD.png}},
J. M. Cordes$^6$\href{https://orcid.org/0000-0002-4049-1882}{\includegraphics[scale=0.08]{ORCIDiD.png}},
M. Cruces$^{7}$\href{https://orcid.org/0000-0001-6804-6513}{\includegraphics[scale=0.08]{ORCIDiD.png}},
S. Dai$^{15,4,1}$\href{https://orcid.org/0000-0002-9618-2499}{\includegraphics[scale=0.08]{ORCIDiD.png}},
V. Gajjar$^8$\href{https://orcid.org/0000-0002-8604-106X}{\includegraphics[scale=0.08]{ORCIDiD.png}},
G. Hobbs$^{4}$\href{https://orcid.org/0000-0003-1502-100X}{\includegraphics[scale=0.08]{ORCIDiD.png}},
C. Jin$^{1}$,
M. Kramer$^{7}$\href{https://orcid.org/0000-0002-4175-2271}{\includegraphics[scale=0.08]{ORCIDiD.png}},
D. R. Lorimer$^{9,10}$\href{https://orcid.org/0000-0003-1301-966X}{\includegraphics[scale=0.08]{ORCIDiD.png}},
C. C. Miao$^{2,1}$\href{https://orcid.org/0000-0002-9441-2190}{\includegraphics[scale=0.08]{ORCIDiD.png}},
C. H. Niu$^{1}$\href{https://orcid.org/0000-0001-6651-7799}{\includegraphics[scale=0.08]{ORCIDiD.png}},
J. R. Niu$^{2,1}$\href{https://orcid.org/0000-0001-8065-4191}{\includegraphics[scale=0.08]{ORCIDiD.png}},
Z. C. Pan$^{1}$\href{https://orcid.org/0000-0001-7771-2864}{\includegraphics[scale=0.08]{ORCIDiD.png}},
L. Qian$^{1}$\href{https://orcid.org/0000-0003-0597-0957}{\includegraphics[scale=0.08]{ORCIDiD.png}},
L. Spitler$^{7}$\href{https://orcid.org/0000-0002-3775-8291}{\includegraphics[scale=0.08]{ORCIDiD.png}},
D. Werthimer$^8$,
G. Q. Zhang$^{11}$,
F. Y. Wang$^{11,12}$\href{https://orcid.org/0000-0003-4157-7714}{\includegraphics[scale=0.08]{ORCIDiD.png}},
X. Y. Xie$^{13}$,
Y. L. Yue$^{1}$\href{https://orcid.org/0000-0003-4415-2148}{\includegraphics[scale=0.08]{ORCIDiD.png}},
L. Zhang$^{14,1}$\href{https://orcid.org/0000-0001-8539-4237}{\includegraphics[scale=0.08]{ORCIDiD.png}},
Q. J. Zhi$^{13,16}$,
Y. Zhu$^{1}$
}

\makeatletter
\def\thanks#1{\protected@xdef\@thanks{\@thanks
        \protect\footnotetext{#1}}}
\makeatother

\begin{document}
\maketitle
\footnotetext[1]{These authors contributed equally to this work.}

\begin{affiliations}

\item CAS Key Laboratory of FAST, NAOC, Chinese Academy of Sciences, Beijing 100101, China
\item University of Chinese Academy of Sciences, Beijing 100049, China
\item Department of Physics and Astronomy, University of Nevada, Las Vegas, Las Vegas, NV 89154, USA
\item CSIRO Astronomy and Space Science, PO Box 76, Epping, NSW 1710, Australia
\item Department of Physics, Anhui Normal University, Wuhu, Anhui 241002, China
\item Cornell Center for Astrophysics and Planetary Science and Department of Astronomy, Cornell University, Ithaca, NY 14853, USA
\item Max-Planck-Institut für Radioastronomie, Auf dem Hügel 69, D-53121 Bonn, Germany
\item Department of Astronomy, University of California Berkeley, Berkeley, CA 94720
\item Department of Physics and Astronomy, West Virginia University, P.O. Box 6315, Morgantown, WV 26506, USA
\item Center for Gravitational Waves and Cosmology, West Virginia University, Chestnut Ridge Research Building, Morgantown, WV, USA
\item School of Astronomy and Space Science, Nanjing University, Nanjing 210093, China
\item Key Laboratory of Modern Astronomy and Astrophysics (Nanjing University), Ministry of
    Education, Nanjing 210093, China
\item Guizhou Normal University, Guiyang 550001, China
\item School of Physics and Technology, Wuhan University, Wuhan 430072, China
\item Western Sydney University, Locked Bag 1797, Penrith, NSW 2751, Australia
\item Guizhou Provincial Key Laboratory of Radio Astronomy and Data Processing, Guizhou Normal University, Guiyang 550001, China
\end{affiliations}

\vspace{0.2in}

\begin{abstract}
The event rate,  energy distribution, and  time-domain behaviour of repeating fast radio bursts (FRBs) contains essential information regarding their physical nature and central engine, which are as yet unknown\cite{petroff19, cordes19}. As the first precisely-localized source, FRB 121102\cite{spitler16,chatterjee17,tendulkar17} has been extensively observed and shows non-Poisson clustering of bursts over time and a power-law energy distribution\cite{michilli18,zhangy18,gourdji19}. However, the extent of the energy distribution towards the fainter end was not known. Here we report the detection of 1652 independent bursts with a peak burst rate of 122~hr$^{-1}$, in 59.5 hours spanning 47 days. A peak in the isotropic equivalent energy distribution is found to be $\sim 4.8\times 10^{37} \ {\rm erg}$ at 1.25~GHz, below which the
detection of  bursts is suppressed. The burst energy distribution is bimodal, and well characterized by a combination of a log-normal function and a generalized Cauchy function. 
The large number of bursts in hour-long spans allow sensitive periodicity searches between 1~ms and 1000~s. The non-detection of any periodicity or quasi-periodicity poses challenges for models involving a single rotating compact object. The high burst rate also implies that FRBs must be generated with a high radiative efficiency, disfavoring emission mechanisms with large energy requirements or contrived triggering conditions. 
\end{abstract}

\begin{figure}[!htp]
\centering
\includegraphics[width=0.9\textwidth]{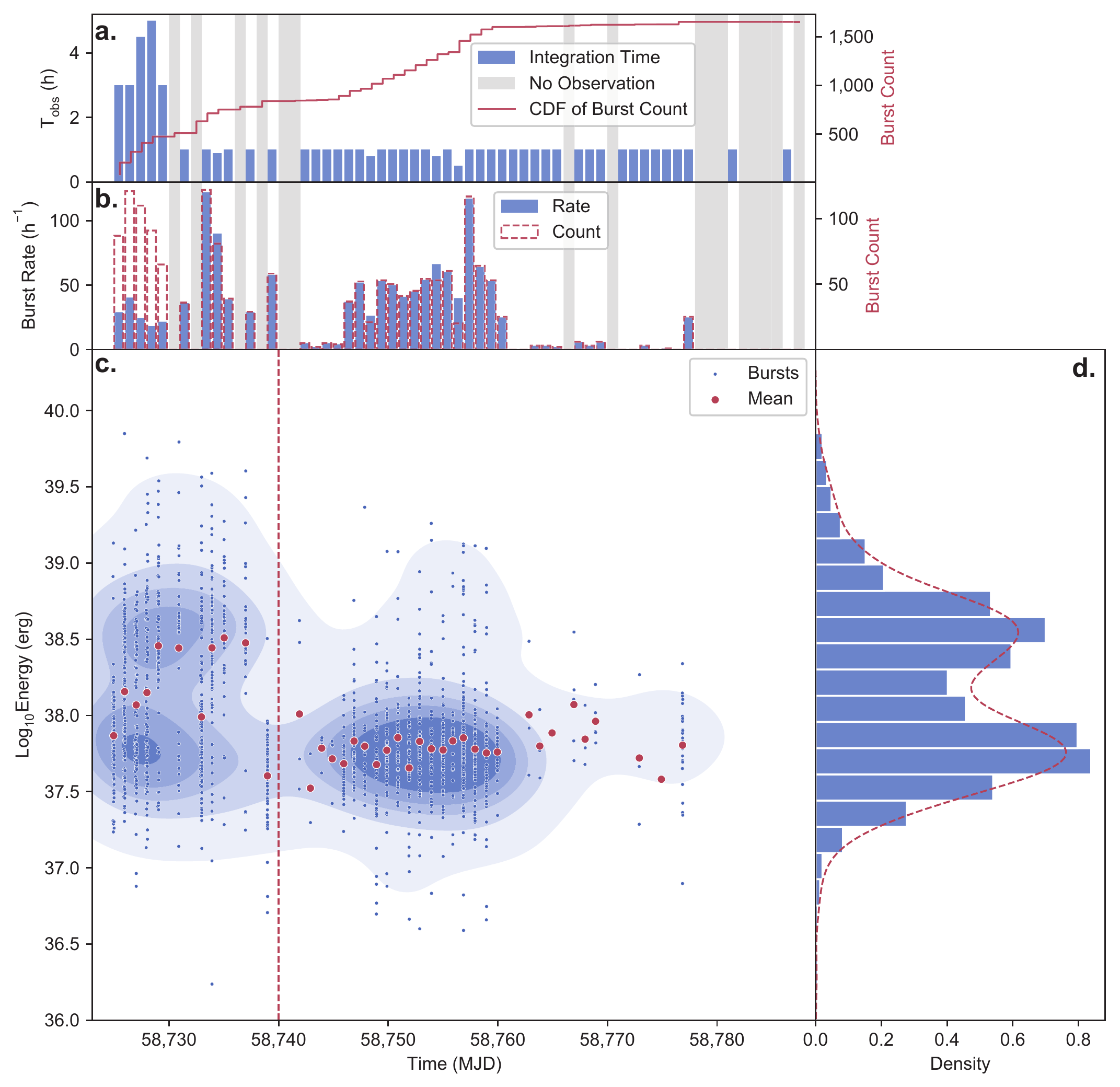}
\caption{{\bf The detected bursts and the temporal energy distribution during the observing campaign.} Panel a: the duration of each observing session (blue bar) and the cumulative number distribution of the bursts (red solid line). Panel b: the rate (blue bar) and count (red dashed line) of the bursts detected. The gray shaded bars denote days without observations. Panel c: Time dependent burst energy distribution. The blue dots are all the 1652 bursts, the red dots represent the average value for each observing session. The blue contour is the 2D Kernel Density Estimation (KDE) of the bursts. Panel d: the isotropic energy histogram of the bursts detected before MJD 58740 (14th Sept. 2019); the red dashed line represents the KDE of this distribution.}
\label{fig:pulse}
\end{figure}

We have been carrying out a continuous monitoring campaign of FRB~121102 with  the Five-hundred-meter Aperture Spherical radio Telescope (FAST\cite{li18}) since August 2019. Between August 29 and October 29, 2019, we detected 1652 independent burst events (see Supplementary Table 1) in a total of 59.5 hours, covering 1.05 GHz to 1.45 GHz with 98.304 $\mu$s sampling and 0.122 MHz frequency resolution. The total number of previously published bursts from this source was 347 (Ref. \cite{zhangy18, gourdji19}
and http://www.frbcat.org). 
The flux limit of this burst sample is at least three times lower than those of previous observations. The cadence and depth of the observations allow for a statistical study of the repeating bursts, revealing several previously unseen characteristics. Fig.~\ref{fig:pulse} depicts 
the burst statistics as a function of time, with the accumulated counts in one-hour bins in the upper panel and the day-to-day average burst rates in the  panel below it. 
 The burst rate peaked at 122~hr$^{-1}$ on September 7th and then 117~hr$^{-1}$ on October 1st, both measured over the respective one-hour session. In both instances, the burst rate dropped precipitously afterwards. Burst energies vs. epoch are shown in the lower left and the energy histogram on the right, showing bimodality that is itself a function of epoch.

\begin{figure}[!htp]
\centering
\includegraphics[width=15cm]{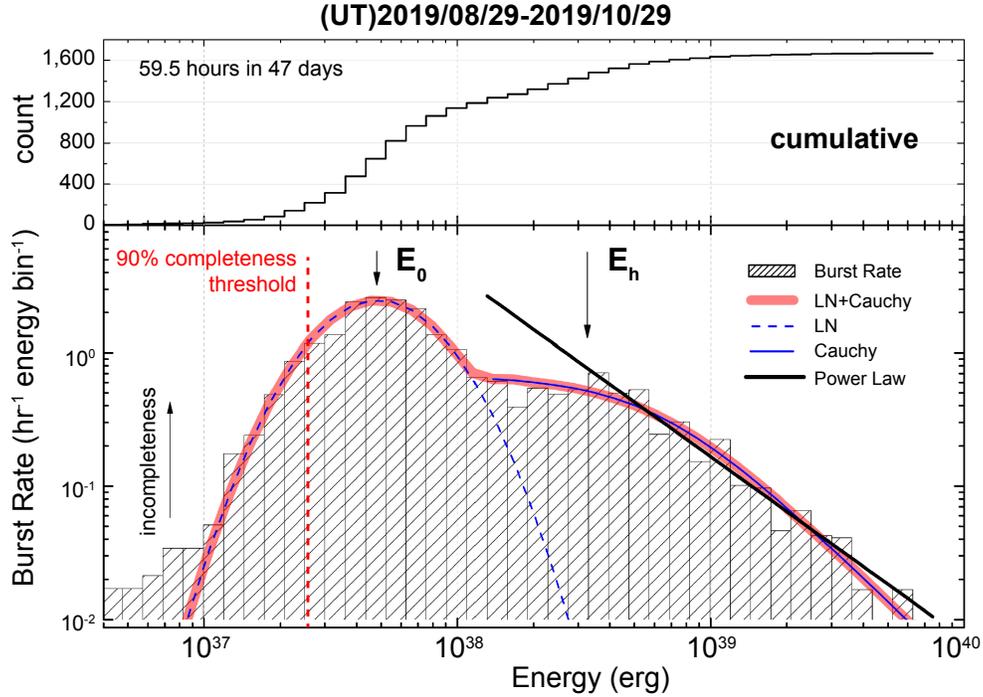}
\caption{{\bf The burst rate distribution of the isotropic equivalent energy at 1.25~GHz for FRB 121102 bursts.} The bimodal `lognormal(LN-dashed blue) + Cauchy (solid blue)' distribution is shown in red and a single power-law fit for bursts above a certain threshold $E \geq E_{h} = 3\times10^{38}$~erg is shown in black.  The 90 percent detection completeness threshold is shown by the red dashed line,  corresponding to $E_{90}$ = $2.5\times10^{37}$~erg for an assumed pulse width of 3~ms and scaling as the square root of the pulse width} (see Fig.~9). The missed weak bursts below $E_{90}$, as indicated by the upward arrow, will make the log-normal distribution wider, but will not affect the location of the peak $E_0$ (Methods).
\label{fig:rate}
\end{figure}

We measured the peak flux density, pulse width, and fluence for each burst. Given the redshift $z=0.193$\cite{tendulkar17}, we adopted the corresponding luminosity distance $D_{\rm L}=949$~Mpc based on the latest cosmological parameters measured by the Planck team\cite{Planck} and calculated the isotropic equivalent energy of each burst at 1.25~GHz (see Methods). The derived energies span more than three orders of magnitude, from below $10^{37}$~erg to near $10^{40}$~erg. Fig.~\ref{fig:rate} presents the histogram of the bursts (lower panel) and the cumulative counts as a function of energy (upper panel). With a prominent peak and two broad bumps, the distribution cannot be fit by a single power-law or a single log-normal function (Table 1). A satisfactory fit can be achieved with a log-normal distribution plus a generalized Cauchy function:
\begin{equation}
N(E) = \frac{N_0}{\sqrt{2\pi}\sigma_E E} \exp\left[\frac{-(\log{E}-\log{E_0})^2}{2\sigma_E^2}\right] +
\frac{\epsilon_{E}}{1+(E / E_0)^{\alpha_E}},
\end{equation}
where $\epsilon_{E} = 0$ for $E<10^{38}$ erg and $\epsilon_{E} = 1$ for $E>10^{38}$ erg. 
The characteristic energy $E_0$ is $4.8\times 10^{37}$~erg and is robust against uncertainties in detection threshold and choices of pipelines (Methods). The best-fit distribution index is $\alpha_E = 1.85\pm0.30$. At high energies, a simple power law consistent with the results derived from previous bursts\cite{gourdji19,shannon18}, can describe the distribution reasonably well. 
It is clear that no single functional form can fit the data in the full energy range. At the low energy end, the log-normal distribution is consistent with a stochastic process. At the high energy end, the generalized Cauchy function describes a steepening power law with an asymptotic slope of $\alpha_E$. Mathematically, a Cauchy distribution, also known as a Cauchy–Lorentz distribution, also describes the ratio of two independent, normally distributed random variables. The bimodality of the energy distribution is also  time dependent, with the high energy mode having more events before MJD 58740, which also help rule out the bimodality being an artifact due to  significant drifts in system calibration over time. 

Each pulse time of arrival (ToA) was transformed to the solar system barycentre using the DE405 ephemeris. No periodicity between 1 ms to 1000 s could be found in the power spectrum calculated using either the Phase-Folding or the Lomb-Scargle Periodogram (see Methods). 

\begin{figure}[!htp]
\centering
\includegraphics[width=13cm]{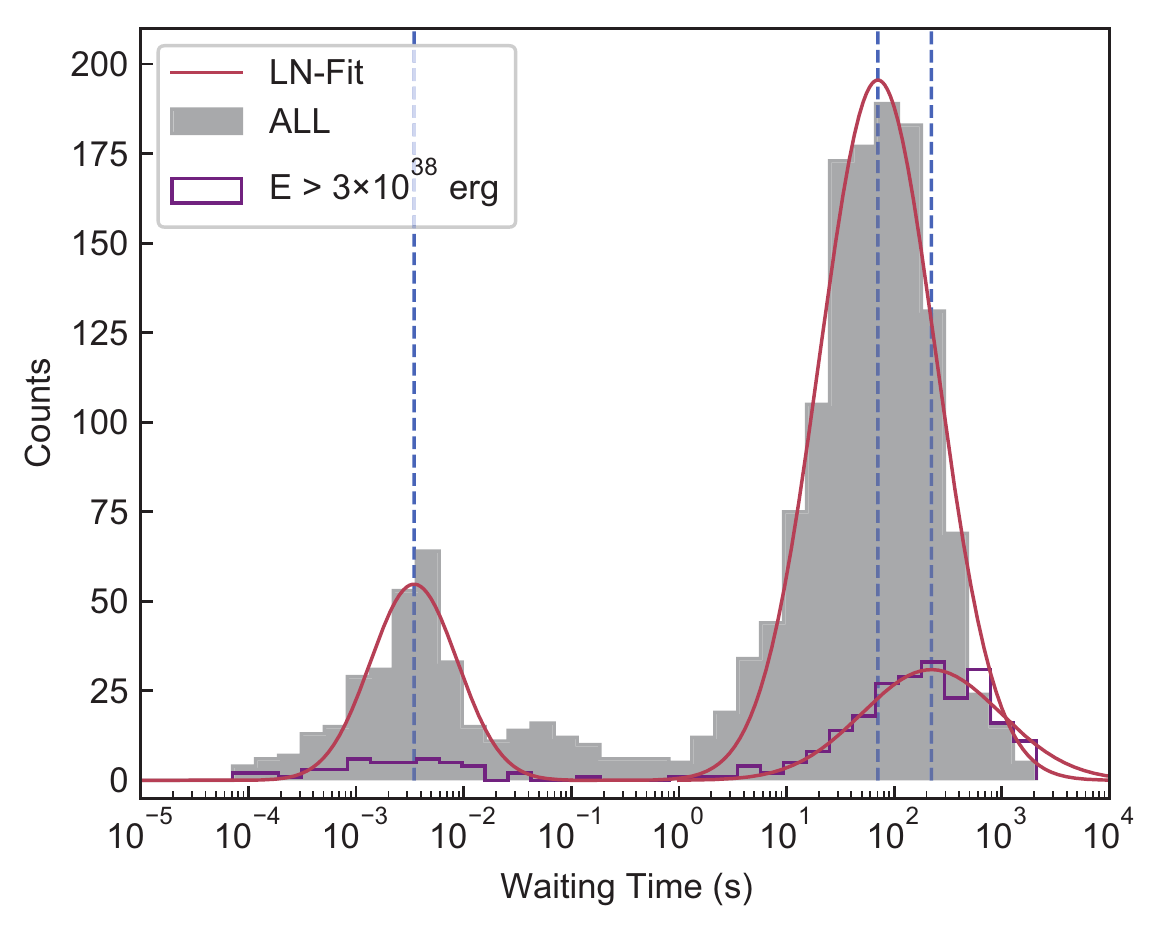}
\caption{{\bf The waiting time distribution of the bursts.} The grey bar and solid red curve denote the distribution of waiting time and its log-normal (LN) fit. The high energy component ($E>3\times10^{38}$~erg) is shown as the solid purple line . The three fitted peak waiting times (blue dashed vertical lines) from left to right are $3.4\pm1.0$~ms, $70 \pm 12$~s, and $220\pm100$~s, respectively. 
The peaks around 70 s and 220 s in the waiting time distribution are close to the  average values for the respective samples (full and high energy). This is consistent with a stochastic process (see main text and Method for further discussion.}
\label{fig:wt_fit}
\end{figure}

The waiting time between two adjacent (detected) bursts is $\delta$t = $t_{\rm i+1}-t_{\rm i}$, where $t_{\rm i+1}$ and $t_{\rm i}$ are the arrival times for the ($i+1$)th and ($i$)th bursts, respectively. 
All waiting times were calculated for pulses within the same session to avoid long gaps of $\sim 24$ h. The distribution of the waiting times (Fig.~\ref{fig:wt_fit}) has a dominant feature that can be well fit by a log-normal function centered at 70$\pm$12s. Selecting only high energy pulses $E > 3\times10^{38}$ erg, the peak moves to 220 seconds.
The waiting time distribution and absence of periodicity are
 generally consistent with previous findings for FRB 121102\cite{katz18,zhangy18,palaniswamy18,gourdji19} and can be reproduced within the uncertainties by simulating bursts arriving randomly in time. For example, the location of the peak of the log-normal distribution can be obtained with a Monte-Carlo simulation mimicking the sampling  cadence and number of detections of the real observations (see Methods). The peaks around 70 s and 220 s in the waiting time distribution are close to the  average values for the respective samples (full and high energy). This is consistent with the waiting time distribution being a combination of a stochastic process and the lack of sampling for time scales longer than $\sim$ 1000s. The secondary peak centered at $\sim 3.4$~ms, however, 
 is most likely due to substructure of individual bursts through some may be closely spaced, independent bursts.

The waiting time distribution and absence of a periodicity are in sharp contrast to expectations from standard radio pulsars, which involve stable rotation and emission in narrow beams from a narrow range of altitudes.  If FRB~121102 involves a rotating object, the periodicity can be erased if beam directions and altitudes are sufficiently stochastic, introducing scatter in arrival times and reducing any features in the power spectrum or waiting time distribution that would signify periodicity. Nonetheless, the 70s waiting-time peak still places an upper bound on the underlying period.

\begin{figure}[!htp]
\centering
\includegraphics[width=17cm]{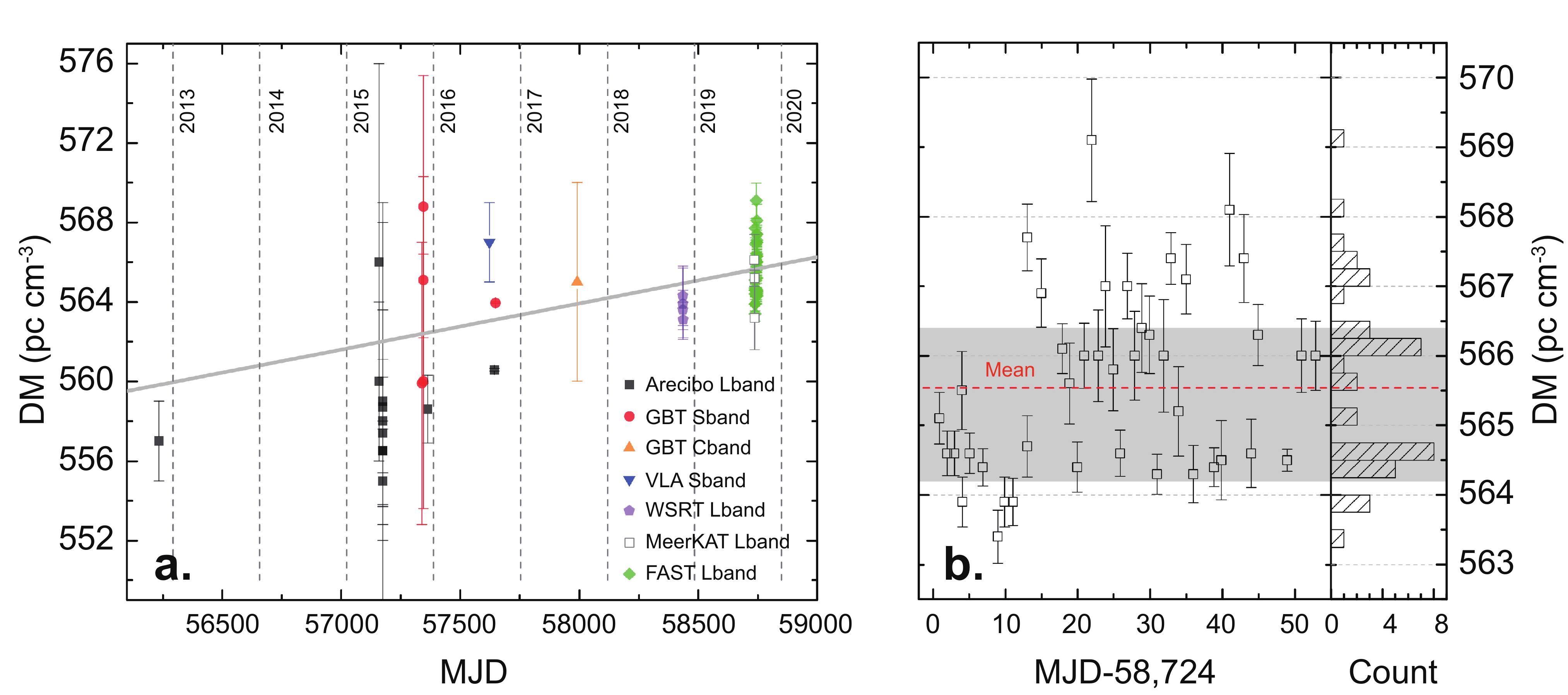}
\caption{{\bf The DM evolution of FRB 121102.} Left panel (a): Temporal DM variation for FRB~121102 over a nine-year period. The solid grey line denotes the  linear fit with  slope $+0.85\pm0.10$~pc~cm$^{-3}$~yr$^{-1}$. Right panel (b): the distribution of  DM estimates for the brightest burst on each day of FAST observations. The red dashed line indicates the average DM of 565.6~pc~cm$^{-3}$ during the FAST observations, and the grey region shows the $95\%$ confidence level.}
\label{fig:DM_time}
\end{figure}

The optimal dispersion measure (DM)  of the bursts is constrained to 565.8$\pm$0.9 ${\rm pc\ cm^{-3}}$ 
between MJD 58,724 and MJD 58,776 (see Methods). This suggests that the DM of FRB 121102 has increased by $\sim 5–8 \ {\rm pc\ cm^{-3}}$ (or $\sim 1.0–1.4\%$) compared to earlier detections\cite{scholz16,petroff16,Oostrum20}, confirming a trend seen before\cite{hessels19}  with larger significance level (Extended Data Fig. \ref{fig:DM_time}). Combining all the data, the averge slope is
\begin{equation}
    \frac{d {\rm DM}}{dt} = +0.85\pm0.10\ {\rm pc\ cm^{-3}\ yr^{-1}}.
\end{equation}
Note that the long term trend relies heavily on earlier measurements, which is further explored in Method.
This is inconsistent with the decreasing-trend predicted for a freely expanding shell (e.g. a supernova remnant) around the FRB source\cite{metzger17}, but is consistent with such a shell during the deceleration (Sedov-Taylor) phase\cite{yangzhang17}. 

We detected no polarization in the bursts at 1.4~GHz, in contrast with higher frequency observations\cite{michilli18} but consistent with previous results at similar frequencies\cite{hessels19} (Methods).

The large sample of bursts sheds new lights on theoretical models of FRBs. The isotropic equivalent energy distribution (or energy function) necessitates a bimodal fit, suggesting possibly more than one emission mechanism or emission site/beam-shape.  The log-normal distribution characterizes the weaker bursts, the generation of which may become less efficient below the characteristic energy scale of $E_0 \sim 4.8\times 10^{37}$ ergs. Some magnetar models do predict a luminosity lower bound for producing FRBs\cite{wadiasingh20,lu20}, and the reported $E_0$ value may be interpreted by adjusting parameters within these models.

As shown in Fig.~\ref{fig:wt_fit}, the distribution of the time intervals between bursts (henceforth referred to as the waiting times) is log-normal in form. This behaviour is similar to that observed in other astrophysical bursting events such as soft gamma-ray repeaters (SGRs)\cite{gogus00,wangyu17}. The extremely-high burst rate revealed by our observational campaign poses challenges to some models invoking an expensive trigger to each burst\cite{Bagchi17,smallwood19}. Models invoking giant-pulse-like emission\cite{cordes16}, active magnetar emission\cite{wadiasingh20,lu20}, or persistent magnetosphere interactions\cite{zhang20} are attractive options but require  masking of the rotational periodicity by stochastic beaming, by large variations in emission altitude, or by propagation delays.

The frequent triggers of bursts also constrain coherent radiation models. In particular, the popular synchrotron maser models demand well-ordered magnetic field lines in the upstream of the shock \cite{metzger19,beloborodov19}. Clustered FRBs require successive shocks propagating into the previously shocked medium, which is hot and probably with distorted magnetic field lines. The short waiting times therefore challenge these models regarding whether coherent emission can be emitted with such short waiting times. 

The synchrotron maser mechanism is also very inefficient\cite{metzger19,beloborodov19}. The total isotropic energy emitted in the 1652 bursts reported in this paper is $3.4 \times 10^{41}$ erg.  Adopting a typical radiative efficiency, the total isotropic energy output during our 47-day observational campaign is already $\sim 37.6\%$ of the available magnetar energy. Considering beaming would not change this estimate significantly (Methods). 
Conversely, coherent emission mechanisms that invoke a neutron star magnetosphere\cite{kumar17,lu20} can radiate in the radio band much more efficiently, and therefore, are preferred by the data.

\bibliographystyle{naturemag}

\begin{addendum}
 \item This work is supported by National Natural Science Foundation of China (NSFC) Programs No. 11988101, No. 11725313, No. 11690024, No.12041303, No. U1731238, No. U2031117, No. U1831131, No. U1831207; by CAS International Partnership Program No. 114-A11KYSB20160008; by CAS Strategic Priority Research Program No. XDB23000000; and the National Key R\&D Program of China (No. 2017YFA0402600); and the National SKA Program of China No. 2020SKA0120200, the Cultivation Project for FAST Scientific Payoff and Research Achievement of CAMS-CAS.
 S.C. and J.M.C. acknowledge support from the National Science Foundation (AAG~1815242).
D.R.L. acknowledges support from support from the National Science Foundation awards AAG-1616042, OIA-1458952 and PHY-1430284.
PW acknowledges support by the Youth Innovation Promotion Association CAS (id.~2021055) and CAS Project for Young Scientists in Basic Reasearch (grant~YSBR-006), PW and CHN acknowledge support from cultivation project for FAST scientific payoff and research achievement of CAMS-CAS. QJZ is supported by the Technology Fund of Guizhou Province ((2016)-4008). LQ is supported by the Youth Innovation Promotion Association of CAS (id.~2018075).
This work made use of data from FAST, a Chinese national mega-science facility built and operated by the National Astronomical Observatories, Chinese Academy of Sciences. 
\item[Author Contributions] 
DL, RD, WWZ launched the FAST campaign; PW, CHN, YKZ, YF, NYT, JRN, CCM, LZ processed the data; DL, BZ, PW drafted the paper; RD, XXZ, VG, CJ, YZ, DW, YLY built the FAST FRB backend; LQ, GH, XYX, QJZ, SD made key contributions to the overall FAST data processing pipelines; LS, MC, MK provided salient information on FRB 121102 from other observatories, particularly Effelsberg, and contributed to the scientific analysis; SC, JMC, DRL, FYW contribued to the writing and analysis, including simulations of the waiting time distribution (JMC). FYW and GQZ made contributions of the time-dependent of bimodal energy distribution.
  
\item[Competing Interests] The authors declare that they have no competing financial interests.

\end{addendum}

\clearpage

\setlength{\tabcolsep}{3mm}{
\renewcommand\arraystretch{0.6}
\begin{longtable}{c c c c c}%
\caption{\bf The fitted parameters of the isotropic equivalent energy distribution. \\
\label{tab:fittab}}
\\
\hline%
\hline%
Function & Fitting parameter & Energy range (erg) && R$^{2}$\ \ $^{\dag}$\\%
\hline%
\endhead%
\hline%
\endfoot%
\hline%
\endlastfoot%
Power law & $\gamma = -0.61 \pm 0.04$ &4$\times$10$^{36}$$\leq E \leq$8$\times$10$^{39}$ && 0.104(6)\\%
& $\gamma = -1.37 \pm 0.18$ &3$\times$10$^{38}$$\leq E \leq$8$\times$10$^{39}$ && 0.999(1)\\%
\hline%
&  $E_{0}$= 7.62$\times$10$^{37}$ (erg)&&&\\%
Lognormal & $N_{0}$= 2.20$\times$10$^{38}$ &4$\times$10$^{36}$$\leq E \leq$8$\times$10$^{39}$ && 0.85(8)\\%
& $\sigma_E$=0.54 &&&\\%
\hline%
Cauchy & $E_{0}$= 8.16$\times$10$^{38}$ (erg)& 4$\times$10$^{36}$$\leq E \leq$8$\times$10$^{39}$ && 0.075(1)\\%
& $\alpha_E$ = 3.02$\pm$0.5 &&&\\%
\hline%

&  $E_{0}$= 7.2$\times$10$^{37}$ (erg)&&&\\%
Lognormal+Cauchy &$N_{0}$= 2.06$\times$10$^{38}$ &4$\times$10$^{36}$$\leq E \leq$8$\times$10$^{39}$ && 0.925(8)\\%
& $\sigma_E$=0.52 &&&\\%
& $\alpha_E$ = 1.85$\pm$0.3 &&&
\end{longtable}}
\begin{tablenotes}
\footnotesize{
\item[$^{*}$] $^{*}$ Uncertainties in parentheses refer to the last quoted digit.
\item[\dag] $\dag$ Coefficient of determination. $R^{2}=1-S_({\rm res}/S_{\rm tot})*[(n-1)/(n-p-1)]$, where $S_{\rm tot}$ is total sum of squares from data, $S_{\rm res}$ is the minimum fitting residual sum of squares.}
\end{tablenotes}

\clearpage
\newpage

\begin{methods}
\setcounter{figure}{0}
\captionsetup[figure]{labelfont={bf},labelformat={default},labelsep=period,name={Extended Data Fig.}}

\subsection{Search procedures and burst energetics}

We carried out blind searches  using two separate software packages, namely Heimdall\cite{petroff15} \& Presto\cite{Ransom2001} with the same search parameter space (DM = 400 to 650 pc~cm$^{-3}$, DM step of 0.2 pc~cm$^{-3}$, and threshold $S_{peak} /$Noise~$\ge 7$). Burst candidates resulting from both searches were kept for further inspections.  The dynamic spectra of all candidates in both the DM=0 time series and the de-dispersed time series at 565 pc~cm$^{-3}$ were then made and manually checked to ensure that the survived candidates have a plausible dispersion sweep and to filter out RFI events.

In order to obtain high quality flux density and polarization calibration solutions, a 1K equivalent noise calibration signal was injected before each session, which was used to scale data to $T_{\rm sys}$ units. The bottom panel of Extended Data Fig. \ref{fig:offpulse} shows the off-pulse brightness (mK s) of the first pulse in each session. The standard deviation of off-pulse brightness is constant within $6\%$ for all observations. The variation in the off-pulse level comes mainly from the zenith angle dependence of the telescope gain. Kelvin units were then converted to mJy using the zenith angle-dependent gain curve, provided by the observatory through quasar measurements. The upper panel of Extended Data Fig. \ref{fig:offpulse} shows the Zenith-angle-dependent gain applied for each pulse. The red dots denote the average gain in each day. For most days, the pulses that have brightness  closest to the average value were taken at  zenith angles $<$ 15 degrees, which corresponds to a  stable gain of 16 K/Jy. 

We calculated the isotropic equivalent burst energy, $E$ following Equation (9) of Ref. \cite{zhang18a}:
\begin{equation}
E = (10^{39} {\rm erg})\frac{4\pi}{1+z} \left(\frac{D_L}{10^{28}{\rm cm}}\right)^{2}
\left(\frac{F_{\nu}}{\rm Jy\cdot ms}\right)
\left(\frac{\nu_c}{\rm GHz}\right),
\end{equation}
where $F_{\nu}$ = $S_{\nu}\times W_{\rm eq}$ is the specific fluence in units of ${\rm erg \ cm^{-2} Hz^{-1}}$ or $\rm Jy \cdot ms$, 
$S_{\nu}$ is the peak flux density which has been calibrated with the noise level of the baseline, and then measured the amount of pulsed flux above the baseline, giving the flux measurement for each pulses at a central frequency of $\nu_c$ = 1.25 GHz, $W_{\rm eq}$ is the equivalent burst duration, and the luminosity distance $D_L$ = 949 Mpc  corresponds to a redshift $z$ = 0.193 for FRB 121102\cite{tendulkar17}.

\subsection{Detection Threshold and Completeness}
The combined effect of sometimes bandwidth-limited structure of FRB bursts and RFI, particularly the satellite bands around 1.2 GHz, affects the actual sensitivity. The representative  7-$\sigma$ detection threshold in the FAST campaign is 0.015~Jy-ms assuming a 1~ms wide burst in terms of integrated flux (fluence), which is a few times better than that previously available from Arecibo.

To quantify  the detection completeness, we differentiate two kinds of signal-to-noise ratios (SNR), namely, the peak flux SNR$_p$ and the fluence flux SNR$_f$ (corresponding to integrated flux and in turn energy). Even for the same pulses, SNR$_p$ and SNR$_f$ behave differently (Fig.~\ref{fig:FRBsim} (right panel) and \ref{fig:completeness}).
The  complexities and the deviation from the radiometer equation mainly arise  from two ‘non-Gaussian’ aspects in the detection processes.   1) The detection software (Heimdall, Presto, etc.) uses the pulse peak SNR$_p$ as the initial selection criterion. Then \sout{down-sampling}
approximate matched filtering
(often just a box-car sum) in time is used to maximize the SNR$_f$ by comparing a series of trial-smoothing results with different width in time. The candidates are then subject to visual inspection.   2) The instrumental background is often ‘RFI’-limited (Extended Data Fig. \ref{fig:DynamicSpectraRFI}), particularly when the pulse is weak and/or bandwidth limited (where the  the dispersed pulse covers only a fraction of the passband). For example, a ‘weak’ (peak flux $<$ 7$\sigma$) but more ‘complete’ pulse, with a sweep covering the full band and a larger pulse width is more easily identified through visual inspection.

In order to quantify these effects, we carried out an experiment by adding simulated pulses into real data (Extended Data Fig. \ref{fig:FRBsim} (left panel)). All pulses were generated assuming DM = 565 pc cm$^{-3}$ and a Gaussian pulse profile (in frequency) and then sampled in time and frequency in exactly the same fashion as those of the FAST data. The frequency bandwidth of pulses is sampled from a normal distribution, similar to that of the observed pulses. The pulse width (in time) was sampled from a log-normal distribution, similar to that of the observed FRB bursts. The SNR$_p$ of the injected pulses was defined as the peak flux divided by the measured RMS of the real data.  More than 1,000 mock FRB bursts were injected into 20 minutes of FAST data. We then processed the simulated datasets through the same pipelines and procedures. As shown in Extended Data Fig. \ref{fig:FRBsim} (left panel), the simulated pulses with sufficient SNR$_f$ can be distinguished from RFI and thus detected, while weaker ones cannot. Fig.~\ref{fig:FRBsim} (right panel) shows histograms of the peak flux SNR$_p$ and the integrated flux SNR$_f$. The behaviours were similar to the previously published simulations\cite{Gupta20}.

Both detection and fluence completeness are illustrated in Extended Data Fig. \ref{fig:completeness}, which shows the recovered fraction of injected ``FRB" pulses as a function of burst fluence and duration. For reference, the 90 percent detection completeness threshold for a characteristic width of 3~ms occurs at a fluence of 0.02 Jy ms, roughly corresponding to $E_{90}$ = $2.5\times10^{37}$ erg. At a given fluence, the longer durations tend to be more incomplete. Figure~\ref{fig:completeness} shows that the search has a high detection completeness ($>$95 \%) for burst durations $<$20~ms when the burst fluence is $>$0.06~Jy~ms, which corresponds an energy of $>7.5\times10^{37}$~erg. Below this energy, the survey will begin to miss wider bursts. The current detections are consistent with event rate starting to drop below $E_0$.

We produce a ‘reconstructed’ energy distribution, by adding the missing fraction back into the sample (Extended Data Fig. \ref{fig:missfraction} (a)) based on the simulated recovery rate. The reconstructed energy distribution has a wider log-normal distribution for the low-energy pulses, but the same peak $E_0$. We reaffirm the existence and robustness of the peak location in this simulation.\\\\

\subsection{Energy Distribution}
The histogram of burst energies exhibits two clearly separable bumps, which can be well fit by two log-normal function (Fig.~ \ref{fig:missfraction} (b), top panel).

The burst rate versus burst energy distribution shows a broad bump centered around $4.8 \times 10^{37}$~erg and a power-law-like distribution with a slope close to $-2$ for the high energy tail. We first fit the distribution with a single power-law, Cauchy, and log-normal function, respectively. The $\chi^{2}$ and R$^{2}$ tests in Table \ref{tab:fittab} show that these single-component models cannot adequately describe the data.

We then test the hypothesis that the energy distribution can be described by a power-law function $N(E) = N_1 E^{-\alpha_{E}}$ in certain energy range $E_i\ \leq E \leq\ E_f$, i.e.
\begin{equation}
N_1 = \frac{N_{ev}(1-\alpha_{E})}{E_f^{1-\alpha_E}-E_{i}^{1-\alpha_E}},
\end{equation}
where $N_1$ is normalization constant, and $N_{ev}$ is the total number of bursts included. For $3\times 10^{38} \leq E \leq 8\times 10^{39}$ erg, the energy function is consistent with a power law with 
$\alpha_E = 1.85\pm0.3$.
Previous works also found power-law energy distributions for the ASKAP sample and all the bursts from the FRB catalog\cite{wangyu17,shannon18,gourdji19,luo18,wang19,lu19,luo20}. 
A bimodal distribution is clearly needed to properly cover the full energy range.
A log-normal function plus a Cauchy function for the high energy range can achieve a satisfactory fit, with a coefficient of determination $R^2 = 0.928$.  

Giant pulses from the Crab pulsar show a power-law amplitude distribution combined with a long tail of ``supergiant pulses'' \cite{cordes04} due to  underlying magnetospheric physics that might elucidate that of FRB~121102. However, the Crab's giant pulses are manifestly tied to the spin of the neutron star whereas the bursts discussed here are not periodic.  Thus a mechanism is needed to erase the periodicity in detected bursts; this could be caused by large variations in altitude or beaming that is not strictly tied to the rotating neutron star.  It could also be caused by propagation effects in the circumsource medium. 

The low-energy component is  similar to distributions seen in many radio pulsars. At the high-energy end, the trend toward a power-law with slope $\alpha_E$ reminiscent of giant pulses from the Crab pulsar. The bimodality of the energy distribution is also clearly time dependent, with the high energy mode having more events before MJD 58740. The fewer high-energy events after that epoch may signify that the bimodality is due to time-dependent lensing but more plausibly might be an analog of ``mode changes'' commonly seen in long period radio pulsars where pulse components change their relative amplitudes and occurrence rates. Further observations can distinguish between these possibilities.

In Fig.~\ref{fig:pulse}, temporal variations are seen for the collective behaviour of each session. There are days with significantly brighter averages, although weak bursts are always present.

\subsection{Analysis with a modified Cauchy function}

A common Cauchy distribution is
\begin{equation}
p\left(x\right)=\frac{1}{\pi\left(x^2+1\right)}
\end{equation}
Cauchy distribution can also be obtained from the distribution of the ratio of two independent normally distributed random variables with zero mean.

Assume that X and Y are independent of each other and obey a normal distribution respectively:
\begin{equation}
f_{X|Y\left(x\right)}=\frac{1}{\sqrt{2\pi}}e^{-x^2/2}
\end{equation}

Then the probability density of Z=X/Y is:
\begin{equation}\begin{split}
f_{X/Y}\left(z\right)&=\int_{-\infty}^{\infty}\left|y\right|f\left(y,\ yz\right)dy
=\int_{-\infty}^{\infty}\left|y\right|f\left(y\right)f\left(yz\right)dy
=\frac{1}{2\pi}\int_{-\infty}^{\infty}{\left|y\right|e^{-y^2/2}}e^{-y^2z^2/2}dy\\
&=\frac{1}{\pi}\int_{0}^{\infty}y\ e^{-\left(z^2+1\right)y^2/2}dy
=\frac{1}{2\pi}\int_{0}^{\infty}e^{-\left(z^2+1\right)u/2}du
=\frac{1}{2\pi}\frac{2}{\left(z^2+1\right)}\\
&=\frac{1}{\pi\left(z^2+1\right)}
\end{split}\end{equation}
As we can see, if we changed the index 2 of the normal distribution to alpha, then the Cauchy distribution will be the general Cauchy function:
\begin{equation}
p\left(x\right)=\frac{1}{\pi\left(x^\alpha+1\right)}
\end{equation}
The generalized Cauchy function can describe the ratio between two normally distributed  variables. Note that the best-fit index of 1.85 is close to 2 with a $\sigma\sim0.3$, which could thus suggest correlated events for generating strong bursts.

\subsection{Width distribution}
Fig.~\ref{fig:width} shows pulse width $W_{\rm eq}$ against flux density (left) and the pulse width distribution of the bursts (right). The equivalent width $W_{\rm eq}$ is defined as the width of a rectangular burst that has the same area as the bursts, with the height of peak flux density denoted as $S_{\rm peak}$. In our sample, several pulses might be described as multiple components in a single burst, if there is ``bridge'' emission (higher than 5 $\sigma$) between pulses for the bursts with a complex time-frequency structure. This results in some bursts having overestimated  equivalent widths. 
The computed equivalent widths range from 0.43~ms to $\sim$40~ms, consistent with a log-normal distribution centered around $\sim$4~ms. This is consistent with the known statistical properties of repeating FRBs \cite{chime-repeaters}.

\subsection{Monte-Carlo simulations of the waiting time distribution}

Following the exact setup of the observations, including starting time, duration, sampling rate, and pulse burst rate, we generate random ToAs through Monte Carlo simulations. The simulation was performed three times, with numbers of generated pulses being $100\times1652\sim1.6e5$, $1\times1652=1652$, and $0.2\times1652\sim330$, respectively.  The distributions of different waiting time sets are shown in Fig.~\ref{fig:MC}. The log-normal distribution appears in the randomly-generated waiting time distribution, and centered at 0.62~s, 61.89~s, and 272.04~s.

The peak waiting time of the log-normal distribution increases as the number of bursts in the simulated sample decreases. Among the 1652 pulses of FRB 121102, 296 have higher energy than $3\times 10^{38}$~erg, which accounts for about one-fifth of the total pulses. The peak waiting time of these 296 pulses is 220~s, which is close to the peak waiting time from the generated $0.2\times1652\sim330$ pulses. The main feature of waiting is the log-normal distribution centered at 70s, which is close to the simulated distribution with 1652 generated pulses.

Our simulations suggest that of the observed log-normal distributions of waiting time centered at 70~s and 220~s, though not an instrumental effect, are nonetheless consistent with emission from a source that emits FRBs randomly or other "masking" factors, such as the rotating attitudes.

\subsection{Periodicity search}

The Lomb-Scargle periodogram (LSP) method\cite{lomb76,scargle82} is widely used to identify periodicities in data that are not uniformly sampled. 

We apply the LSP method to the ToAs of FRB121102 to determine if there is a possible period. If the bursts of FRB121102 do have a period, folding the burst arrival times according to this period would show clustering in burst phase.

Fig.~\ref{fig:LS} shows periodograms of bursts from FRB121102 bursts for periods ranging from 1~ms to 100~d, where the left panel covers periods from 0.01~d to 100~d and the right panel is for periods from 1~ms to 1000~s. Of the five peaks in the left panel, four are at periods of 0.998~d, 0.499~d, 0.333~d, and 0.153~d, corresponding to the daily sampling and its higher harmonics. The fifth peak at 
$\sim 24$~d yields a non-random but broad distribution of burst phases (Fig.~\ref{fig:LS}, bottom left) that most likely reflects non-uniform detections over the 47~d data span. 
The marginally significant peak at 10.575$\pm$0.008~ms in the top-right panel appears to be related to a large multiple of the original 98.304 $\mu s$ sample interval of the data.  Folding with that period does not show any concentration in pulse phase (bottom right).

In addition to a search for a constant period over the 47~d data set, we also searched for 
periods  ($P$) between $1\ ms$ and $1000\ s$ accompanied by a period derivative ($\dot P$) between  $10^{-12}$ and $10^{-2}\ s\,s^{-1}$ and the same negative $\dot P$ range to fold all the pulses.  This also  did not reveal any underlying  period. Additionally, we divided the pulses according to energy with dividing lines at $5\times10^{37}\ {\rm erg}$ and $3\times10^{38}\ {\rm erg}$, and found that the pulses in different energy intervals do not have significant periodicity. All observing session fall in the predicted active phase of FRB 121102\cite{rajwade20}, thus the addition of this sample does not alter the 157d period found there.

\subsection{DM variation}

Before we can study the detailed emission characteristics of the bursts, the optimum dispersion measure (DM) should be determined. De-dispersed pulse profiles were created for each DM trial between 500 to 650~pc~cm$^{-3}$ with a step size of 0.05~pc~cm$^{-3}$, using the single pulse search tools in Presto \cite{Ransom2001}. Gaussians (multiple when necessary) were fitted to the profiles. The derivative of each Gaussian was then squared. For multiple components, the squared profiles were summed.
The optimum DM was then identified according to the maximization of the area under the squared derivative profiles, thus maximizing the structures in the frequency integrated burst profile. The typical DM optimization method and de-dispersed profiles are demonstrated in Fig.~\ref{fig:DM_profile}.

The resulting histogram distribution of DMs are shown in Fig.~\ref{fig:DM_distribution} (left panel), the optimal value is 565.8$\pm$0.3$\pm$0.76~pc~cm$^{-3}$, between MJD 58,724 and MJD 58,776. The two uncertainties are statistical error and systematic error. The latter is estimated by measuring the $\Delta$DM that results in a DM time delay across the whole band equal to half the equivalent width of the bursts (the typical value is 1.5 ms). This suggests that the DM of FRB 121102 has increased by 5-8~pc~cm$^{-3}$ (1.0-1.4\%) with more than 2.0 $\sigma$ significance compared with earlier detections from MJD 57,364, where Scholz et al. \cite{scholz16} found the optimal value to be 558.6$\pm$0.3$\pm$1.4~pc~cm$^{-3}$ with the similar methodology. Furthermore, the measured DM values and their uncertainties of the bursts are shown in Fig.~\ref{fig:DM_time} as a function of individual observations. DM apparently increased over the last 6 years\cite{petroff16,Oostrum20} and is found to be consistent with a DM growth rate of $+0.85\pm0.10\ \rm pc\ cm^{-3}\ yr^{-1}$ in Equation(2).

For inspecting the reliability of this DM variation trend, we divded the DM measurement into three time bins according to the dates of the event and generated mock DM values in each bin,  based on the mean and the standard deviation of the measured DMs in the respective time bins. The null hypothesis test was then carried out based on the generated DMs under the assumption that DM does not change over time. For each set of generated DM, a slope was fitted. Based on 20,000 trials, the $\sigma$ of the resulting slope distribution is 0.38 $pc\ cm^{-3}\ yr^{-1}$, which is shown in Extended Data Fig. \ref{fig:DM_distribution} (right panel). Thus, the fitted DM growth rate of 0.85 could result from a null hypothesis sample at a 2.6\% probability, slightly better than 2 $\sigma$. This is apparently less significant than the simple fitting, but probably more realistic.

\subsection{Polarization Characteristics}
The polarization was calibrated by correcting for differential gains and phases between the receivers through separate measurements of a noise diode injected at an angle of $45^{\circ}$ from the linear receivers. The circular polarization is consistent with noise, lower than a few percent of the total intensity, which agrees with Ref.\cite{michilli18}. We searched for the rotation measure (RM) from $-6.0\times10^5$ to $6.0\times10^5$\,$\mathrm{rad\,m^{-2}}$, a range that is much larger than the $\mathrm{RM} \sim 10^5\,\mathrm{rad\,m^{-2}}$ reported in Ref.\cite{michilli18}, no significant peak was found in the Faraday spectrum. The linear polarization becomes negligible at L band compared to almost 100\% linear polarization at C band reported in Ref.\cite{michilli18}. For our data, we estimate the depolarization fraction $f_{\rm depol}$ using 
\begin{equation}
\label{eq:depol}
f_{\rm depol} = 1 - \frac{{\rm sin}(\Delta{\theta})}{\Delta{\theta}},
\end{equation}
where the intra-channel Faraday rotation $\Delta{\theta}$ is given by

\begin{equation}
\label{eq:faraday_smearing}
\Delta\theta = \frac{2{\rm RM}_{\rm obs}c^2\Delta\nu}{\nu_c^{3}},
\end{equation}
where $c$ is the speed of light, $\Delta{\nu}$ is the channel width, and $\nu_c$ is the central channel observing frequency. Taking ${\rm RM}_{\rm obs} = 10^5\,\mathrm{rad\,m^{-2}}$ reported in Ref.\cite{michilli18}, $\Delta{\nu} = 0.122\,\mathrm{MHz}$, and $\nu_c = 1.25\,\mathrm{GHz}$ for our data, 
we get $f_{\rm depol} = 20\%$. Although the depolarization fraction is not negligible, the non-detection of the linear polarization cannot be caused by depolarization assuming ${\rm RM}_{\rm obs} < 10^5\,\mathrm{rad\,m^{-2}}$.

We are confident about the non-detection. We have applied the same analysis procedures to bright pulsars and retrieved expected results. The same procedure has been used in multiple publications\cite{feng20,luo20nat,zhangCF20}. Our non-detection does not conflict with the previous almost 100\% linear polarization because all previous polarization detections were accomplished at frequency bands higher than L-band ($\sim$1.4\,GHz). Ref.\cite{michilli18} published polarization measurements at 4-8\,GHz and Ref.\cite{hi21} at 3-8\,GHz. 
Unless the rotation measure was somehow much larger during the FAST observations compared to previous determinations,  the non-detection appears to require strong frequency evolution of the linear polarization.

\subsection{Energy budget constraint on the synchrotron maser magnetar model}
The total isotropic energy emitted in the 1652 bursts reported in this paper is $3.4 \times 10^{41}$ erg. 
We consider that each FRB has a beaming factor of $f_b = \delta\Omega/4\pi < 1$ (where $\delta\Omega$ is the solid angle of the emission of individual burst).
If these individual bursts are isotropically distributed in sky, even though the energy budget for each FRB is smaller by a factor of $f_b$, there would be also approximately $f_b^{-1}$ more undetected bursts (whose emission beams elsewhere) so that the total energy is not changed\cite{zhang20}. Adopting a typical radiative efficiency $\eta \sim 10^{-4} \eta_{-4}$ from numerical simulations\cite{plotnikov19} (which could be even lower in view of the very rapid repetition rate), the total energy emitted solely during the $\sim 60$ hours of observation spanning in 47 days is already $\sim 6.4\times 10^{46} \eta_{-4}^{-1}$ erg (considering FAST only observed $\sim$ 60 hours during these 47 days and assuming that the observed rate applies to the epochs of no observations as well). The total magnetic energy of a magnetar with a surface magnetic field strength $B=10^{15} B_{15}$ G is $\sim (1/6) B^2 R^3 \simeq 1.7 \times 10^{47}$ erg. One can see that the total energy emitted during our observational campaign is already $\sim 37.6\%$ of the available magnetar energy. One possible way to avoid this criticism is to argue that there is a ``global beaming factor'' $F_b = \Delta\Omega / 4\pi$ which is smaller than unity but greater than $f_b$ to describe the beaming angle of all emitted FRBs. This factor could be of the order 0.1 for pulsar-like emission due to the geometry defined by the magnetosphere configuration of the central source. However, for relativistic shocks invoked in the synchrotron maser models, $F_b$ would not be much less than unity due to the lack of a collimation mechanism for a Poynting-flux-dominated outflow.

In view of the fact that FRB 121102 is already active for nearly a decade and that many faint bursts such as the ones reported in this paper have escaped detection from previous telescopes, we believe that the synchrotron maser model is significantly challenged from the energy budget point of view. A previous analysis using data observed by the Green Bank
Telescope\cite{zhangy18} obtained a similar result\cite{wu20}. Even if this global beaming factor $F_b$ = 0.1 is assumed, the released energy during this active period is already $3.8\%$ of the total magnetar energy budget. This disfavors the synchrotron maser model and any model that invokes a low radio radiative efficiency.

\bibliographystyle{naturemag}

\subsection{Data availability}
All relevant data for the 1652 detected burst events are summarized in the manuscript Supplementary Table. Observational data are available from the FAST archive\footnote{http://fast.bao.ac.cn} one year after data-taking, following FAST data policy. Due to the large data volume for these observations, interested users are encouraged to contact the corresponding author to arrange the data transfer.

\subsection{Code availability}
Computational programs for the FRB121102 burst analysis and observations reported here are available at https://github.com/NAOC-pulsar/PeiWang-code. Other standard data reduction packages are available at their respective websites:\\
PRESTO: https://github.com/scottransom/presto\\
HEIMDALL: http://sourceforge.net/projects/heimdall-astro/\\
DSPSR: http://dspsr.sourceforge.net\\
PSRCHIVE: http://psrchive.sourceforge.net

\clearpage

\begin{figure}[!htp]
\centering
\includegraphics[width=14cm]{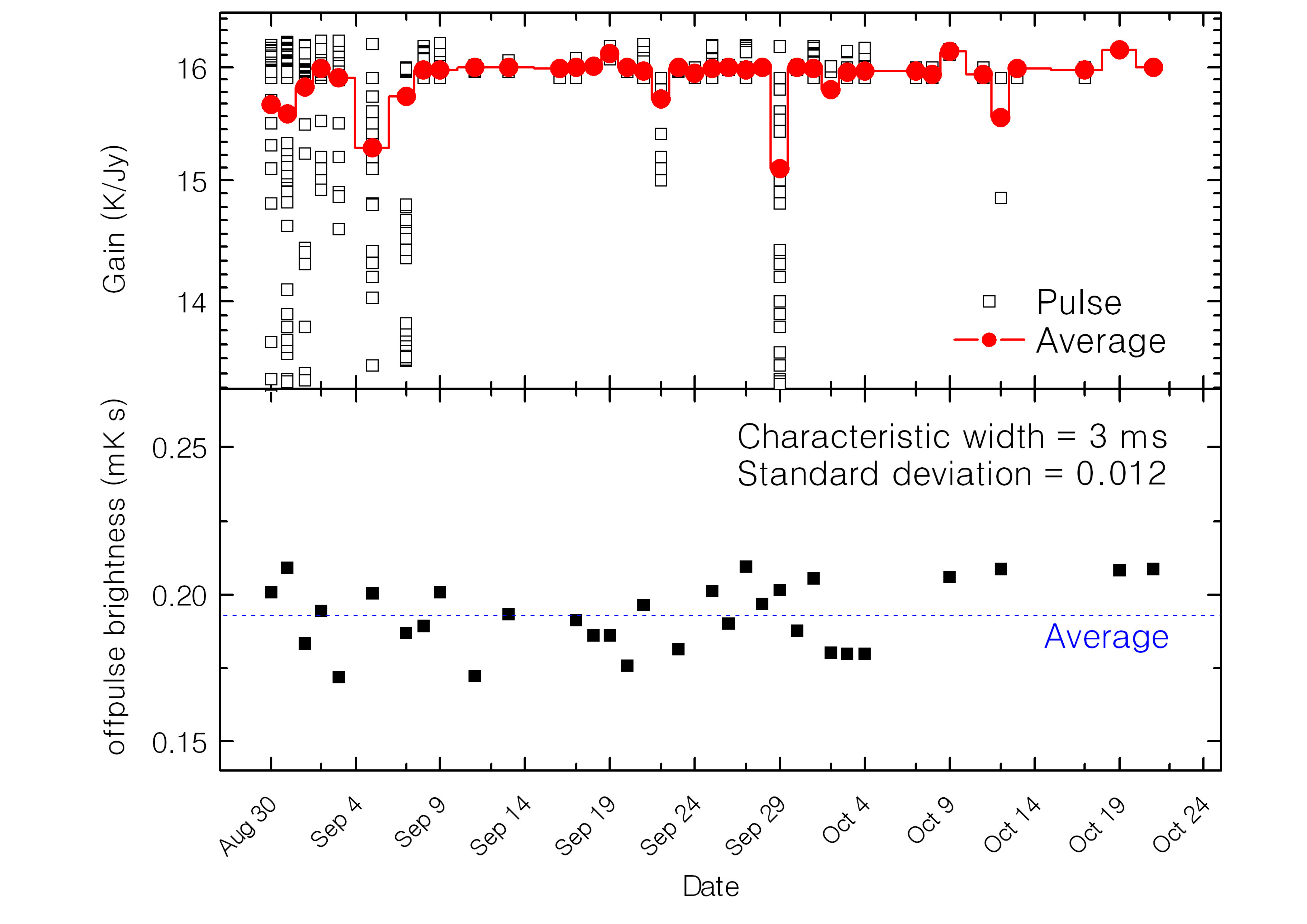}
\caption{{\bf The distribution of the instrumental gain and off-pulse brightness RMS at 1.25~GHz for observations.} The upper panel indicates the gain applied for each pulses. The red dots denote the averaged gain in each day. The bottom panel shows the off-pulse brightness RMS (mK~s) of the first pulse detected each day.}
\label{fig:offpulse}
\end{figure}

\begin{figure}[!htp]
\centering
\includegraphics[width=16cm]{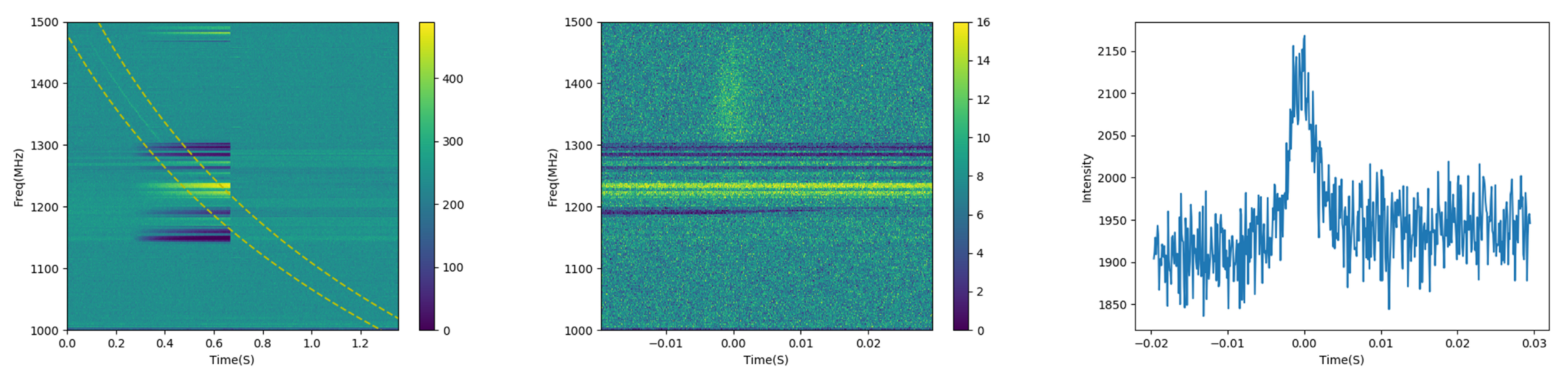}
\caption{{\bf Example of a dynamic spectrum of burst with RFI.}}
\label{fig:DynamicSpectraRFI}
\end{figure}

\begin{figure}[!htp]
\centering
\includegraphics[width=17cm]{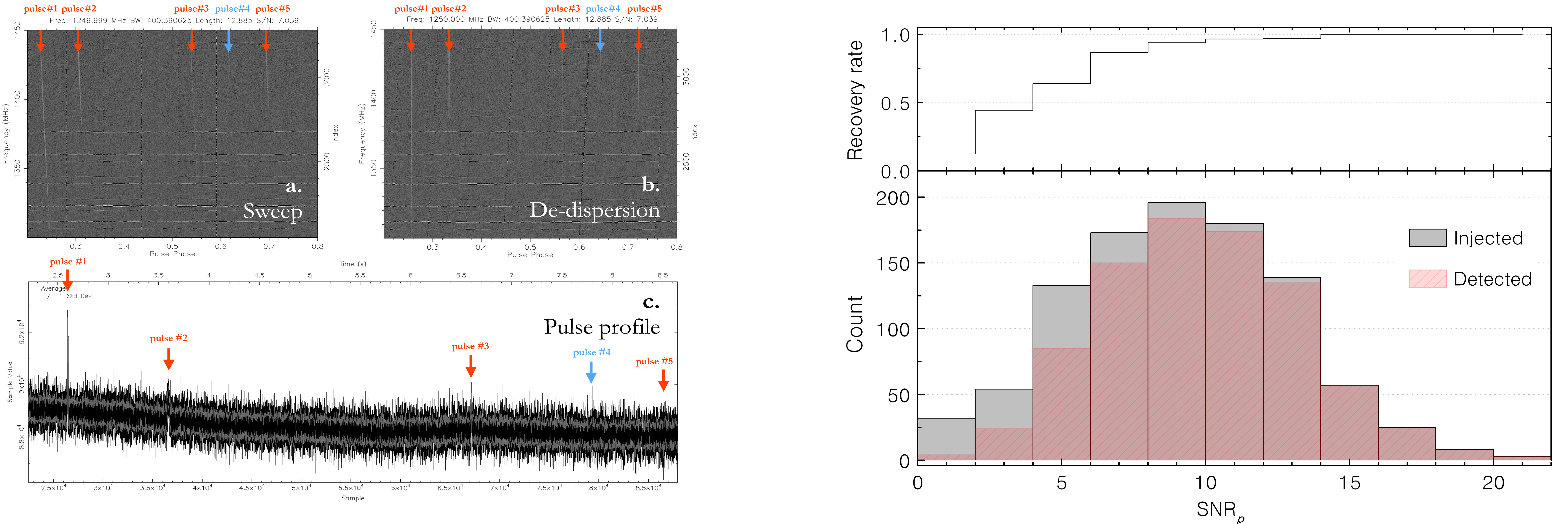}
\caption{{\bf Left panel: Example of FRB simulations.} Upper panels (a and b) are injected and de-dispersed dynamic spectra respectively. The time series is shown in panel (c) with the red arrows pointing to simulated pulses that were detected, while the blue arrow indicates an undetected pulse. 
{\bf Right panel: Comparison of SNR$_p$ recovered by FRB search versus the corresponding injected values.} The SNR$_p$ histograms separately indicate the injected FRB pulses (grey lines) and the mock FRBs detected (red lines).}
\label{fig:FRBsim}
\end{figure}

\begin{figure}[!htp]
\centering
\includegraphics[width=13cm]{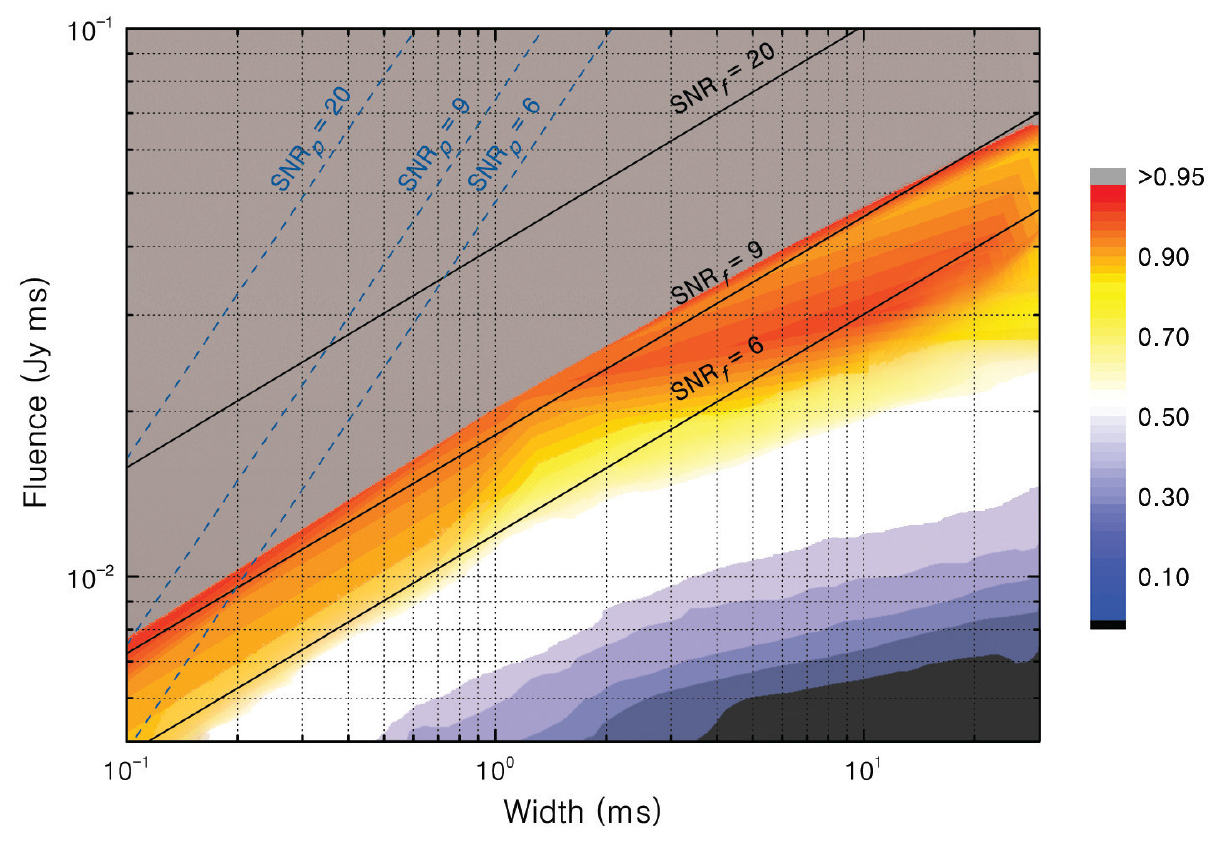}
\caption{{\bf The completeness fraction  of the FAST survey to FRBs as a function of the observed fluence and detected width.} All FRBs lying in the integrated SNR$_f$ $<$6 region are below PRESTO’s search threshold. The region above the integrated SNR$_f$ of 6 shows the incompleteness of our FAST detection to broad FRBs as revealed by the injections. The map was smoothed (rebin) the map with a box of 0.05ms $\times$ 0.002 Jy ms, which ensured the presence of at least one injected pulse in most map areas. Then for a few grid points without pulses, a simple linear interpolation was used to improve the visual appearance.
The colour bar on the right side indicates the detection recovery fraction.}
\label{fig:completeness}
\end{figure}

\begin{figure}[!htp]
\centering
\includegraphics[width=11cm]{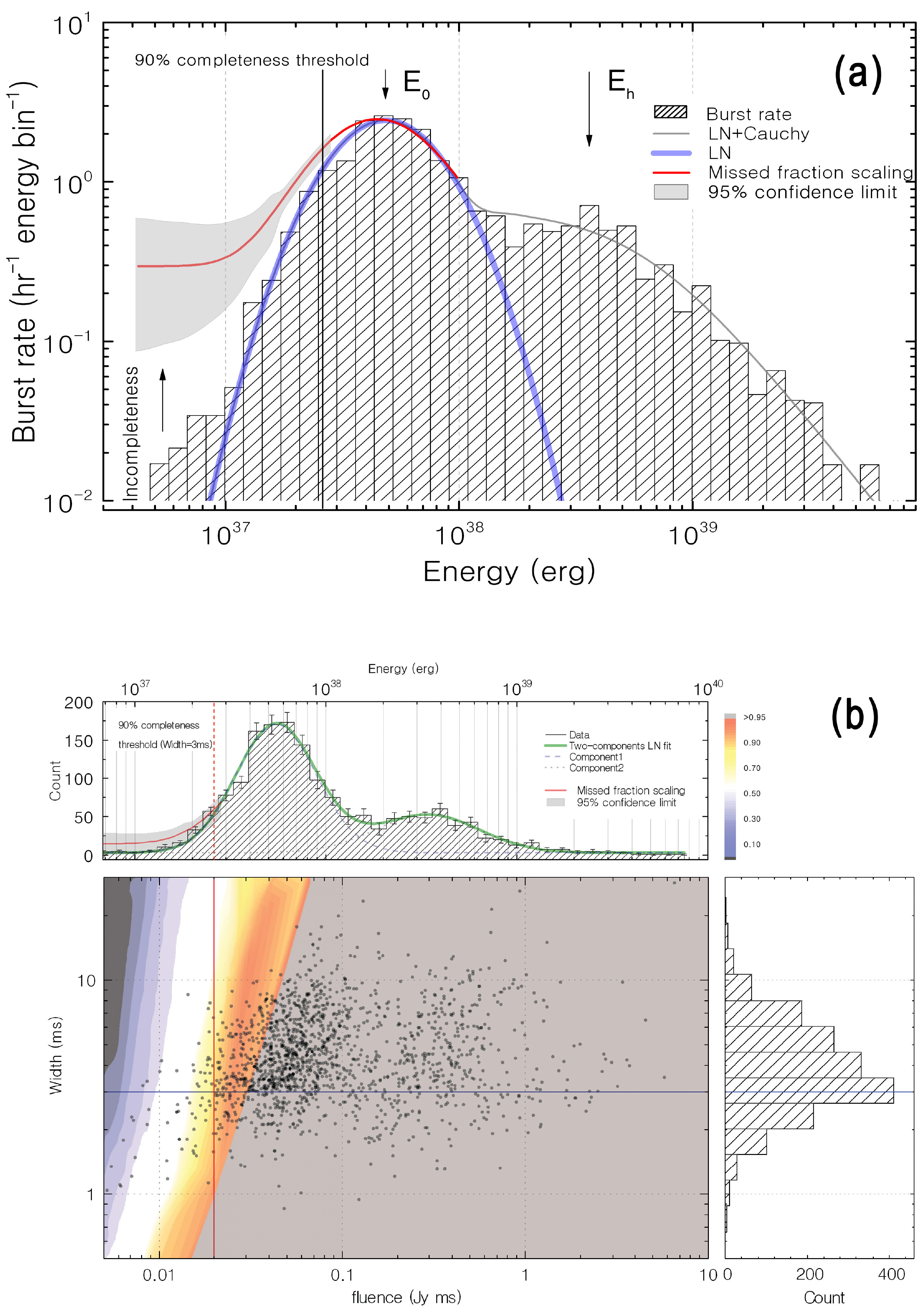}
\caption{{\bf Upper panel (a): The burst rate distribution of the isotropic equivalent energy.} Details as per Extended Data Fig. \ref{fig:rate}. The red line represents the recovered distribution by adding back the missing fraction based on the simulation. The grey shaded region is the uncertainty for a 95$\%$ confidence based on the Poisson statistical assumption in the "reconstructed" fitting.
{\bf Bottom panel (b): The fluence-width distribution at 1.25~GHz for FRB~121102 bursts.} The black dots indicate the  1,652 detected bursts, the colorbar is consistent with Extended Data Fig. \ref{fig:completeness}. In the upper panel, the two-component lognormal (LN) distribution is separately fitted in blue dashed line and grey dot line, an overall fit for bursts is shown in green. The red line and the shaded region indicates reconstructed missing fraction of bursts detection and uncertainty.}
\label{fig:missfraction}
\end{figure}

\begin{figure}[!htp]
\centering
\includegraphics[width=13cm]{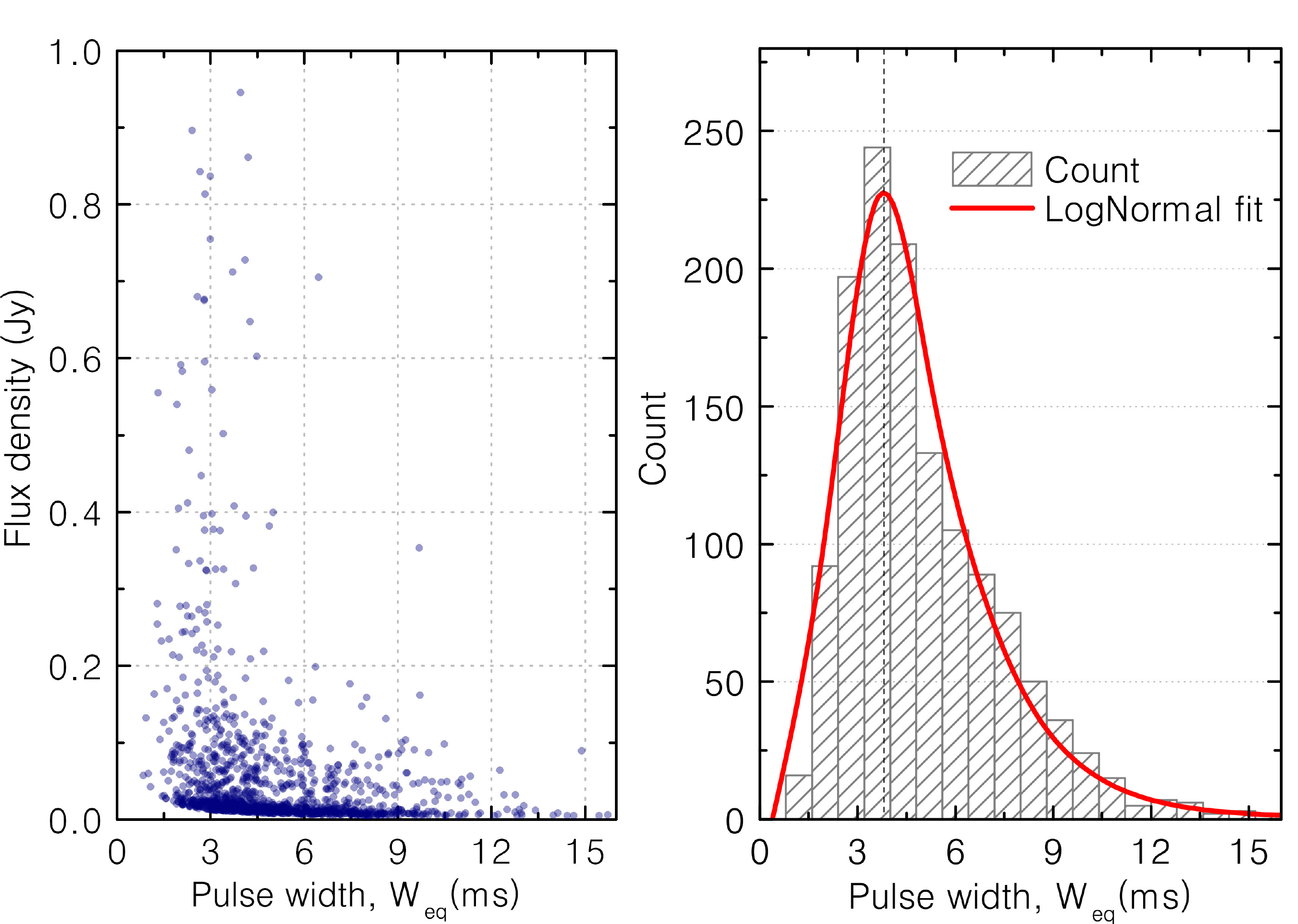}
\caption{{\bf Flux intensity and pulse width distribution of FRB~121102.} {\em Left:} Flux intensity against pulse width for the FRB~121102 bursts with peak SNR$_p$$>$ 10 in our sample. {\em Right:} The equivalent pulse width histogram.}
\label{fig:width}
\end{figure}

\begin{figure}[!htp]
\centering
\includegraphics[width=0.9\textwidth]{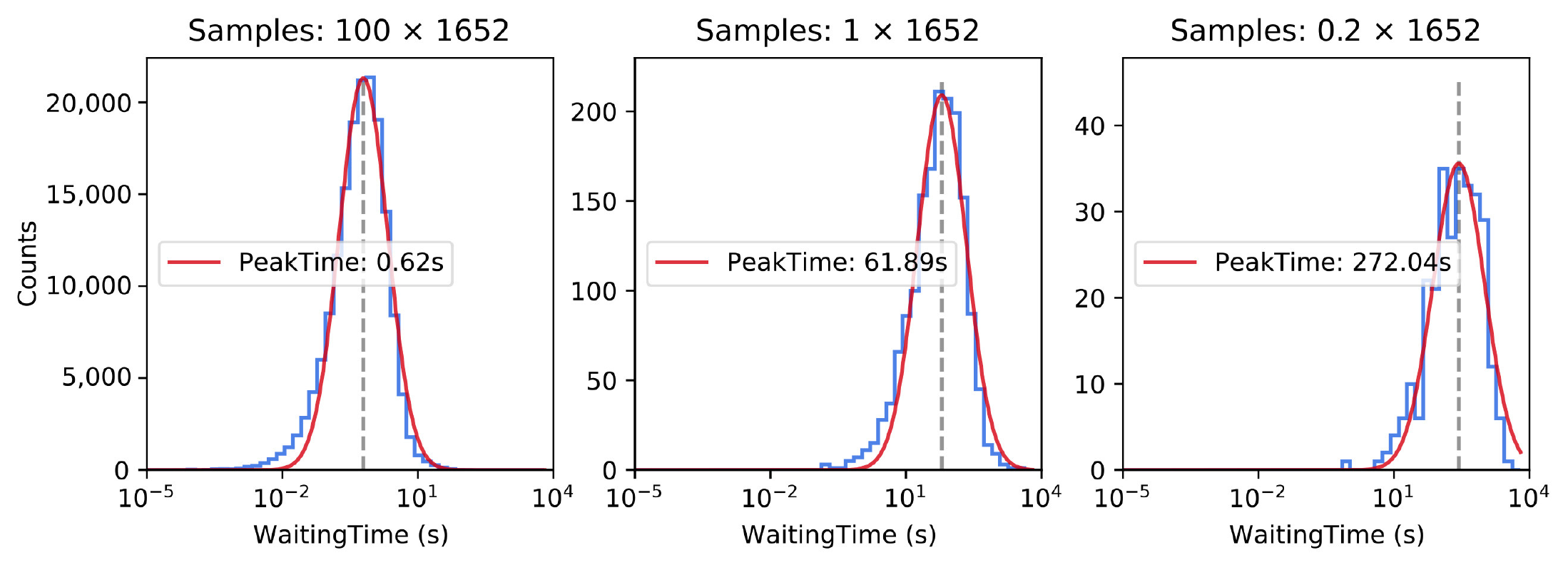}
\caption{{\bf MC simulations of the waiting time distribution.} The three figures correspond to three different simulations, and the number of randomly generated pulses in each simulation are $100\times1652\sim1.6e5$, $1\times1652=1652$, and $0.2\times1652\sim330$. The peak times of the three log-normal distributions are 0.62~s, 61.89~s, and 272.04~s, respectively.}
\label{fig:MC}
\end{figure}

\begin{figure}[!htp]
\centering
\includegraphics[width=0.9\textwidth]{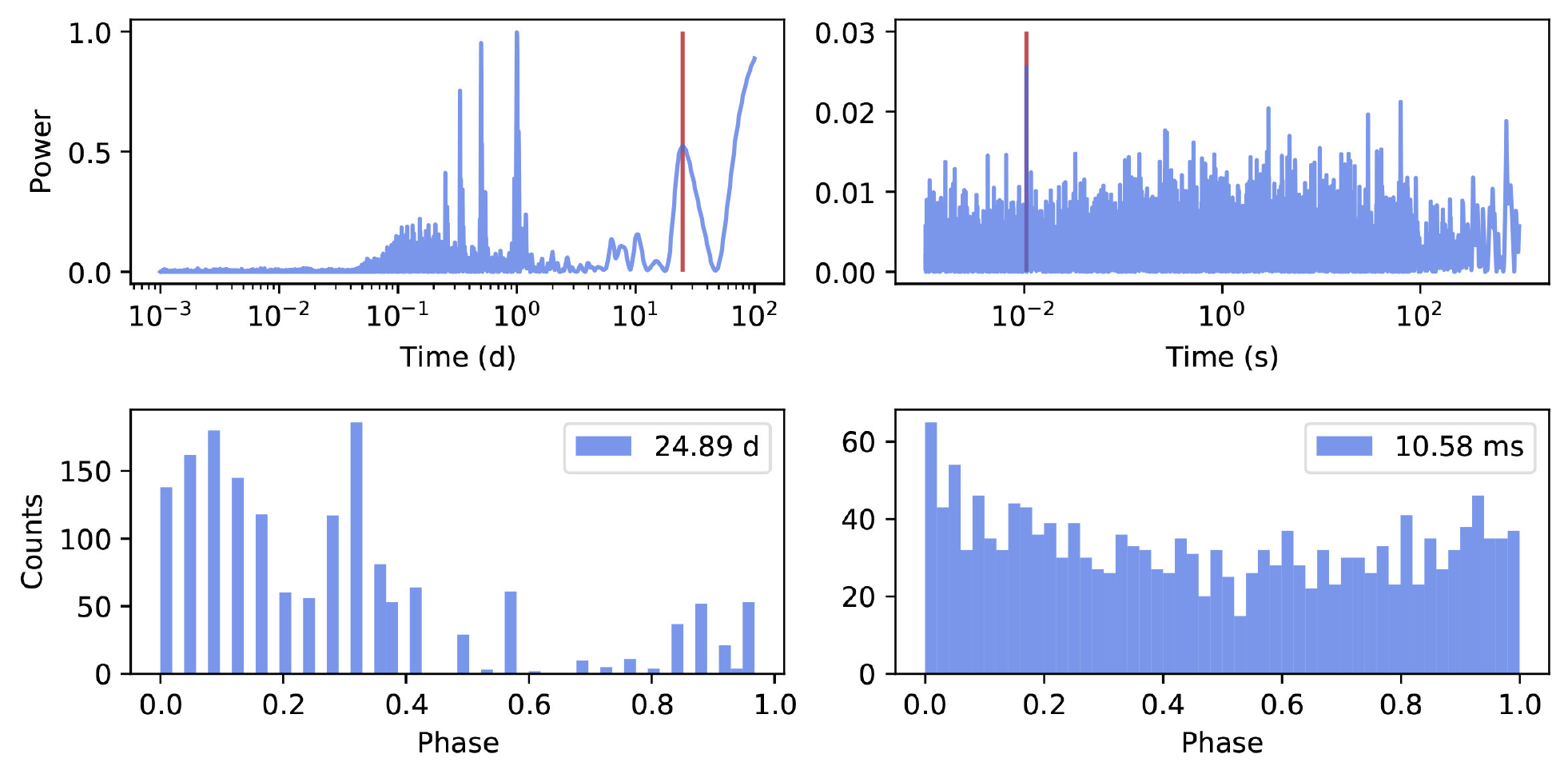}
\caption{{\bf Lomb-Scargle periodograms of FRB~121102 burst arrival times (top row) along with phase histograms for two trial periods (bottom row).} Left: Periods from $10^{-3}$ to $10^2$~d. 
The four leftmost peaks in the periodogram are caused by daily sampling and its harmonics.  The peak at $\sim 24$~d is related to the sampling window function (i.e. non-uniform sampling) over the 47~d data set, as is consistent with the broad distribution in burst phase (bottom left).
Right: Periods from 1~ms to $10^3$~s. The peak at $10$~ms is a large multiple of the original sampling time and also yields no distinct concentration in burst phase.  }
\label{fig:LS}
\end{figure}

\begin{figure}[!htp]
\centering
\includegraphics[width=12cm]{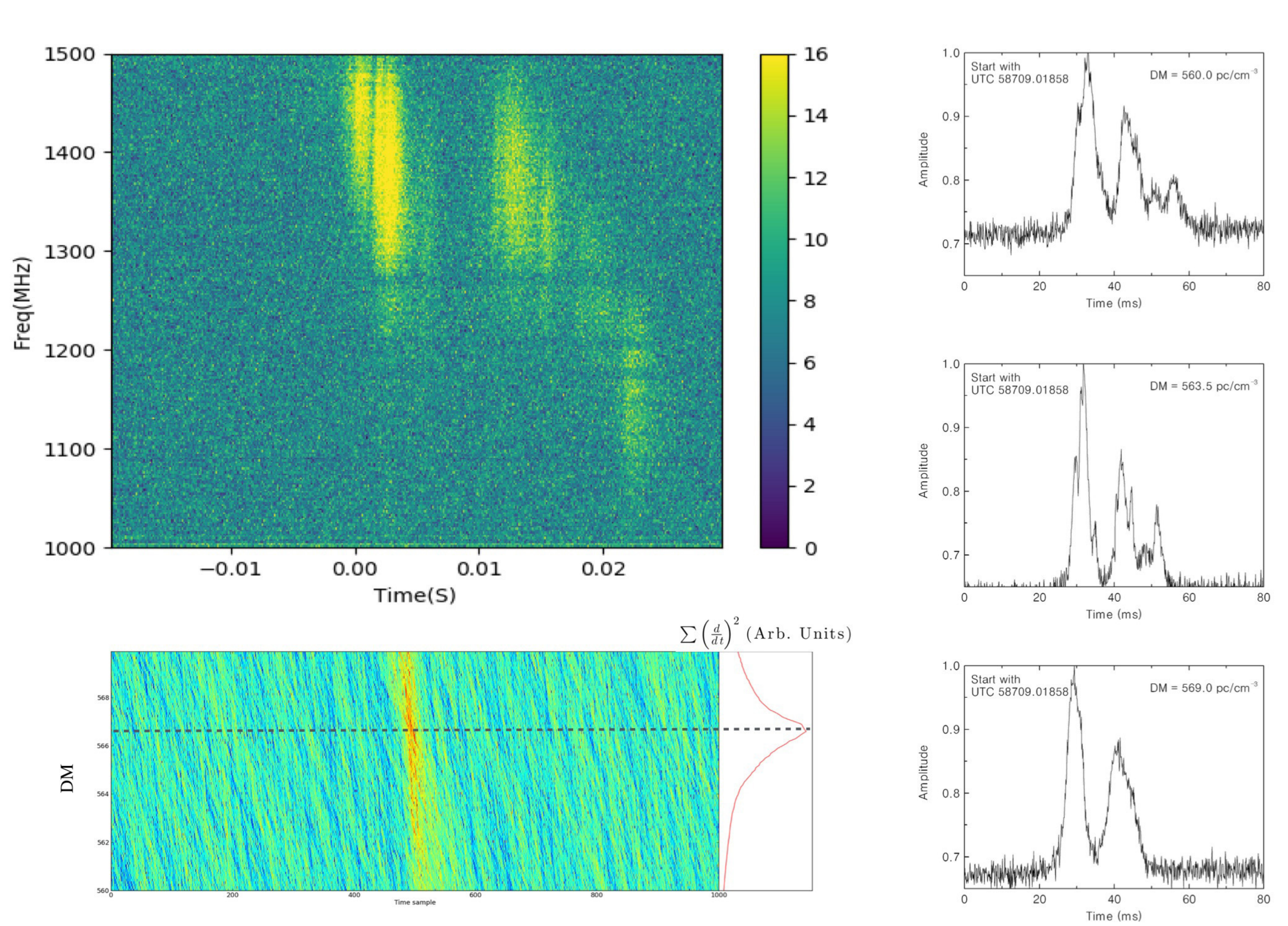}
\caption{{\bf Example of the DM optimization method for FRB~121102.} The complex time–frequency structures for the burst of MJD 58729.01858 was revealed with an optimal DM of 563.5~pc~cm$^{-3}$.}
\label{fig:DM_profile}
\end{figure}

\begin{figure}[!htp]
\centering
\includegraphics[width=16cm]{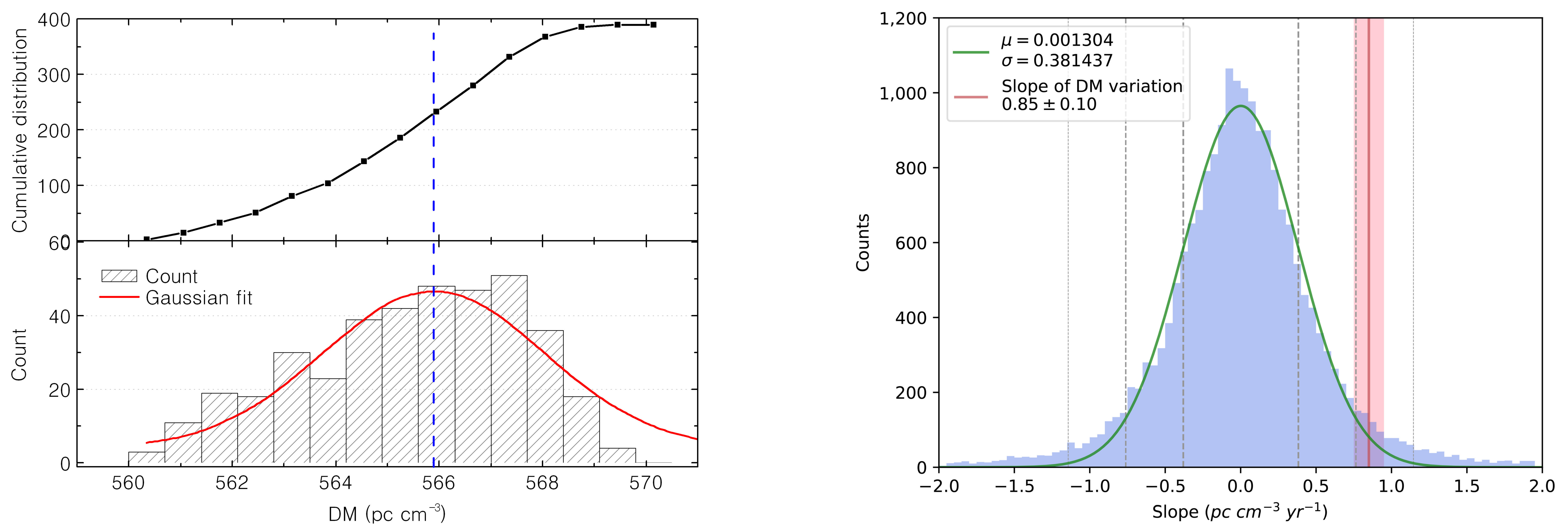}
\caption{{\bf Left panel: Histogram and cumulative distribution of dispersion measure for FRB~121102. Right panel: Slope distribution of null hypothesis test.}}
\label{fig:DM_distribution}
\end{figure}

\end{methods}

\clearpage
\section*{Extended Data}

\renewcommand{\baselinestretch}{1.0}
\selectfont

\noindent
\EXTTAB{tab:bursttab}: {\bf The properties of the 1652 FRB121102 bursts detected by FAST.}

\setcounter{figure}{0}
\setcounter{table}{0}
\captionsetup[table]{name={\bf Supplementary Table}}
\captionsetup[figure]{name={\bf Extended Data Figure}}

\setlength{\tabcolsep}{3mm}{
\renewcommand\arraystretch{0.8}
\scriptsize
\begin{longtable}{c c c c c c c c c c}%
\caption{\bf}
\\
\hline%
\hline%
Burst\ ID&Burst time$^{a)}$&DM&Width&Bandwidth$^{b)}$&Peak\ Flux&Fluence&Energy\\%
&(MJD)&(pc cm$^{-3}$)&(ms)&(GHz)&(mJy)&(Jy\ ms)&(erg)\\%
\hline%
\endhead%
\hline%
\endfoot%
\hline%
\endlastfoot%
1     &  58724.877574724 & 567.3 ( 12 ) & 6.13 ( 9 ) & 0.11 & 12.831 $\pm$ 0.2 & 0.0787 ( 2 ) & 8.426 ( 16 ) $\times 10^{ 37 }$ \\%
2     &  58724.879789116 & 564.3 ( 7 ) & 4.0 ( 3 ) & 0.25 & 28.25 $\pm$ 1.7 & 0.1130 ( 0 ) & 1.210 ( 0 ) $\times 10^{ 38 }$ \\%
3     &  58724.885144007 & 565.1 ( 3 ) & 2.83 ( 0 ) & 0.4 & 269.05 $\pm$ 1.9 & 0.7614 ( 0 ) & 8.151 ( 0 ) $\times 10^{ 38 }$ \\%
4     &  58724.886345270 & 568.3 ( 5 ) & 6.35 ( 9 ) & 0.26 & 7.003 $\pm$ 0.1 & 0.0445 ( 1 ) & 4.764 ( 8 ) $\times 10^{ 37 }$ \\%
5     &  58724.886384602 & 565.8 ( 5 ) & 8.51 ( 9 ) & 0.16 & 6.268 $\pm$ 0.1 & 0.0534 ( 1 ) & 5.711 ( 9 ) $\times 10^{ 37 }$ \\%
6     &  58724.887370014 & 568.8 ( 5 ) & 4.25 ( 6 ) & 0.12 & 4.474 $\pm$ 0.1 & 0.0190 ( 0 ) & 2.036 ( 3 ) $\times 10^{ 37 }$ \\%
7     &  58724.887596868 & 567.6 ( 6 ) & 5.49 ( 7 ) & 0.2 & 12.48 $\pm$ 1.3 & 0.0685 ( 1 ) & 7.334 ( 8 ) $\times 10^{ 37 }$ \\%
8     &  58724.891114144 & 565.8 ( 5 ) & 7.99 ( 9 ) & 0.23 & 7.943 $\pm$ 0.1 & 0.0635 ( 1 ) & 6.796 ( 10 ) $\times 10^{ 37 }$ \\
\end{longtable}}
\begin{tablenotes}
\item[\#] $\#$ Uncertainties in parentheses refer to the last quoted digit.
\item[a)] $a)$ Arrival time of burst peak at the solar system barycenter, after correcting to the frequency of 1.5GHz.
\item[b)] $b)$ A conservative 30$\%$ fractional error is assumed.
\end{tablenotes}

\end{document}